\documentclass[12pt]{iopart}
\usepackage{graphicx}
\usepackage{overpic}
\usepackage{amssymb}
\usepackage{hyperref}

\begin{document}
\title{Organic crystalline polymers: structural properties and way to synthesis under high pressure.}
\author{ V.V. Brazhkin $^1$, N.A. Nikolaev $^1$, Y.M. Shulga $^2$ , Y.B. Lebed $^3$, M.V. Kondrin $^1$}
\address{ $^1$ Institute for High Pressure Physics RAS, 108840 Troitsk, Moscow, Russia}
\address{ $^2$ National University of Science and Technology 'MISIS', Leninsky Pr. 4, 119049 Moscow, Russia}
\address{ $^3$ Institute for Nuclear Research RAS,  Moscow, Russia}
\ead{brazhkin@hppi.troitsk.ru,mkondrin@hppi.troitsk.ru}

\begin{abstract}
We consider different  structures, which can be obtained by polymerization of aromatic organic molecules under high pressures. These 2D and 3D covalently bonded organic polymers  and their functionalization can pave the way to production of energy storage and conversion devices. High-pressure synthesis might serve as a useful hint for production of these structures and their functionalized analogs by means of wet chemical synthesis. 
\end{abstract}

\section{Introduction}
Discovery of graphene stimulated  intensive searches of its analogs and derivatives. One of the most
interesting derivatives is hydrographene called graphane \cite{sluiter:prb03,sofo:prb07}, which, according to calculations \cite{wen:pnas11} is
thermodynamically more favorable than all C-H hydrocarbons including benzene. At the same time,
pressure-induced polymerization of hydrocarbons and their derivatives at room temperature results in
formation of amorphous \cite{cansell:jcp93,ciabini:prb05,ciabini:nm07,citroni:pnas08} or poorly ordered one-dimensional products,such as polyethylene and
benzene-derived carbon nanothreads \cite{fitzgibbons:nm14}.

High-pressure high-temperature synthesis of bulk graphane with  C - H(D) composition
from benzene and pyridine (C - H - N$_{0.2}$ was reported earlier\cite{kondrin:cec17}). Graphane samples were  several millimeter in  size with  relatively large (several nanometer) size of crystalline grains of sp3-bonded graphane lattice, as revealed by X-ray diffraction and  transmission electron microscopy. Most hydrogen atoms in the samples are in  C - H groups connected by aliphatic bonds.
The synthesized graphanes  are stable up to 500 C at atmospheric pressure. The macroscopic density of
CH samples is 1.5 - 1.57 g/cm$^{- 3}$ and the refractive index 1.78 - 1.80.  The absorption spectrum of samples
with a high degree of crystallization  exhibit weak absorption maximum at 2.8 eV, which is responsible
for yellow - orange color, large absorption maximum at 4 eV and absorption edge associated with
the width of the optical gap at 5.2 eV. The bulk modulus (30 - 37 GPa) and shear modulus (15 - 18 GPa)
of the samples fabricated , as well as their hardness (1 - 1.5 GPa), are about twice as high as respective
values of polycrystalline graphite. There were also reports of synthesis of structurally similar products by hydrogenation of graphite powder under high pressure and high temperature conditions \cite{bashkin:jetpl04:en,antonov:c16} and by``wet synthesis'' from fluorographene \cite{yang:c16}.

\begin{figure}
\includegraphics[width = \columnwidth]{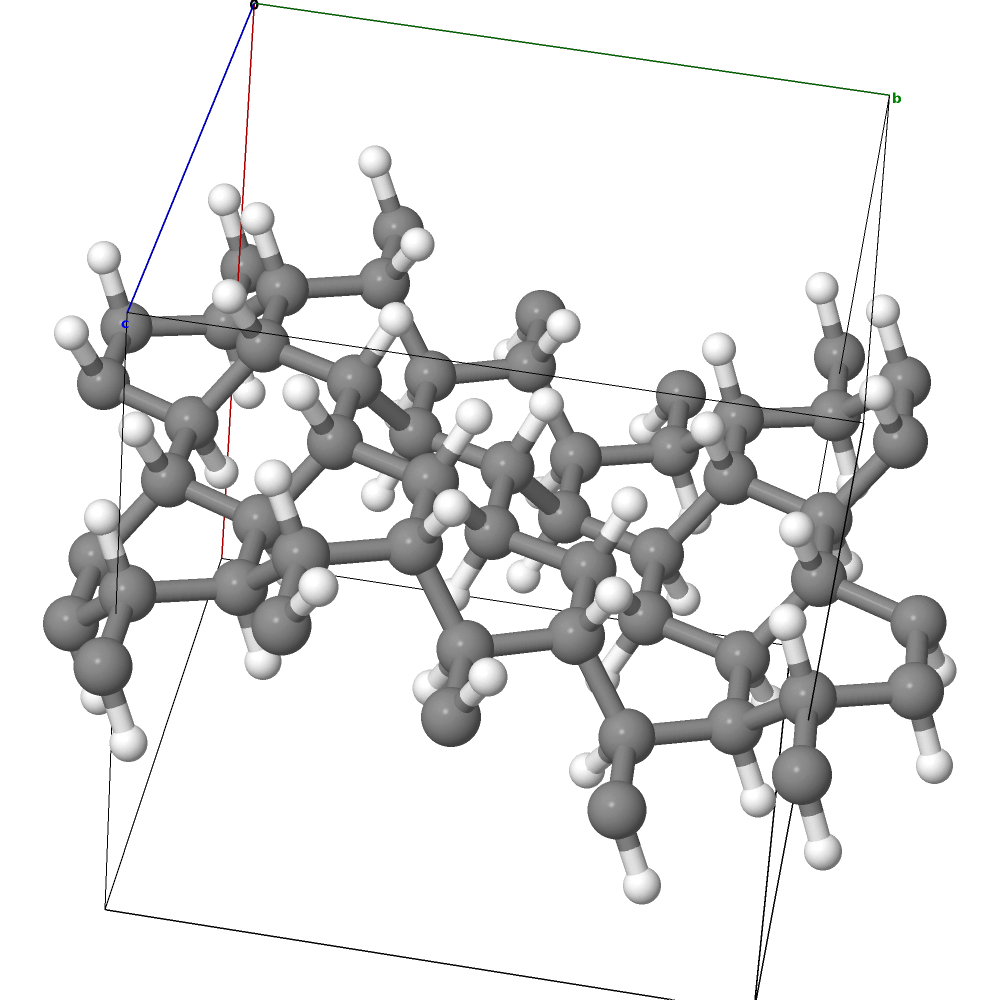}
\caption{Example of hybrid graphane structure (3-cycle-2-step), consisting of cyclohexane rings in chair and boat configurations.}
\label{f0}
\end{figure}

In these works there is  some disagreement about the structure of obtained product, which is augmented by the tendency to regard as ``true'' graphane only one of its chair conformations (so called A-type graphane sheet).  A-type graphane is the least dense and the most energetically favorable conformation at ambient pressure. Obviously, and according to {\em ab-initio} calculations, more compact graphane modifications of should be more stable at high pressures. It was demonstrated, that at high pressure even   boat configurations of graphanes(C and D according to Ref.~\cite{wen:pnas11}) with lonsdaleite type of C-C bond are more energetically favorable than benzene.  Interplay between graphane forms can lead to even more complicated structures, one of which,  a combination of A- and B-graphanes named tri-cycle graphane, was first proposed in Ref.~\cite{he:pssr12}. Subsequently, it was demonstrated that combination of all four simple graphane polymorphs (A-D) is also possible and it was proposed that just this type of hybrid graphane structure does grow under high-pressure high-temperature (HPHT) conditions by polymerization of benzene \cite{kondrin:cec17}. The simplified example of such structure 3-cycle-2-step is shown in Fig.~\ref{f0} \cite{supplementary}. Although this structure (3-cycle-2-step) doesn't match diffraction patterns of polymerized benzene phases obtained under high pressure (they possess significantly larger unit cells), it may serve a simple model  for  the study of various optical modes by first principle calculations. Since calculations of vibrational modes are very demanding to computational resources, the use of simpler structure saves us a lot of time. In this paper, we provide additional evidence for presence of complex hybrid structure of graphanes synthesized under high pressure.

Once the existence of layered polymeric hydrocarbons is established, why not search for completely covalently bound organic structures?  Theoretical efforts in this direction were already undertaken \cite{lian:jcp13,he:jpcm13,lian:sr15,kondrin:pccp15,kondrin:acb16}. To build 3D hydrocarbon structures, one has to deal with steric interactions between hydrogen atoms (free volume is limited by shortness of C-C bond 1.54 \AA\ in comparison to C-H bond 1.1 \AA\ ) and large strains induced by deformation of ideal sp$^3$ bonded carbon network. So far, only one
 family of 3D CH polymorphs \cite{kondrin:acb16} turned out more stable than benzene. Thus, there is hope to synthesize these polymers from it. At this type of synthesis one should consider ``topological'' obstacles, which consists in necessity of breaking benzene's double C=C bonds. In Ref.~\cite{kondrin:acb16}, to bypass these obstacles, copolymerization of benzene with acetylene  was proposed ~\cite{kondrin:acb16}. Experimental results on copolymerization of benzene and acetylene were recently reported \cite{huang17,ward17} and we also discuss them in our paper.

Another  search direction of high-dimensional crystal organic polymers is to vary chemical composition. Substitution of one CH group for nitrogen was realized in experimental studies of pyridine polymerization \cite{kondrin:cec17}, where formation of crystal phase under HPHT conditions was confirmed by X-Ray methods. In this connection,  investigation of  hydrogen cyanide (HCN)  polymerization would be interesting, as a simpler approximation of reaction with pyridine. Polymeric HCN  attracts considerable attention of astronomers as possible component of comets' cores and interstellar matter. There is a rather disputable hypothesis  about polymeric HCN as possible source of life on Earth (as a DNA precursor) \cite{matthews:fd06}. 

In this paper, we mostly rely on the results of DFT calculations and try to compare them with already known experimental results. In any case, we hope that our findings will serve as practical guide for synthesis of proposed materials.
\section{Methods}

In our {\em ab-initio} calculations QuantumESPRESSO software package was used\cite{gianozzi:jopcm09}. For the density functional calculation we employed the Perdew-Burke-Ernzerhof exchange correlation method with norm-conserving pseudopotentials \cite{pp-remark} for all atoms with the energy cutoff 70 Ry and the charge-density cutoff 800 Ry. For integration over Brillouin zone unshifted Monkhorst-Pack grid with granularity finer than $4 \times 4 \times 2$ was used. In the process of calculation, relaxation of cell dimensions and ions' positions (with initial symmetry fixed) was performed, until residual force on every atom did not exceed 0.001 Ry/bohr and additional stress -- 0.5 kbar. Intensities of infrared active modes and phonon dispersion curves were calculated using PHONON program from the QuantumESPRESSO suite. Powder X-Ray diffraction patterns were simulated with FULLPROF package \cite{pbcm:fullprof93,rodriguez:01}. Visualization of crystal structures and their vibrations was done by means of JMol program \cite{jmol}.

\section{Infrared spectra of graphanes}
Experimental data on optical properties of bulk graphane was reported in Refs.~\cite{antonov:c16,yang:c16,kondrin:cec17}. Due to poor crystallization of the samples, strong luminescence band caused by impurities, prevented registration of the Raman signal in Refs.~\cite{antonov:c16,kondrin:cec17}. The Raman spectra provided in Refs.~\cite{elias:s09,yang:c16} strongly resemble those of fluorographite and graphene,  so we believe that the Raman spectra may correspond to unreacted initial components. Below we will restrict our discussion only to IR-spectra. It is worth to mention that all considered polymeric structures have inversion center, so all optically active modes can be either IR- ({\em ungerade}) or Raman ({\em gerade}) active. 

\begin{figure}
\includegraphics[width = 0.9\columnwidth]{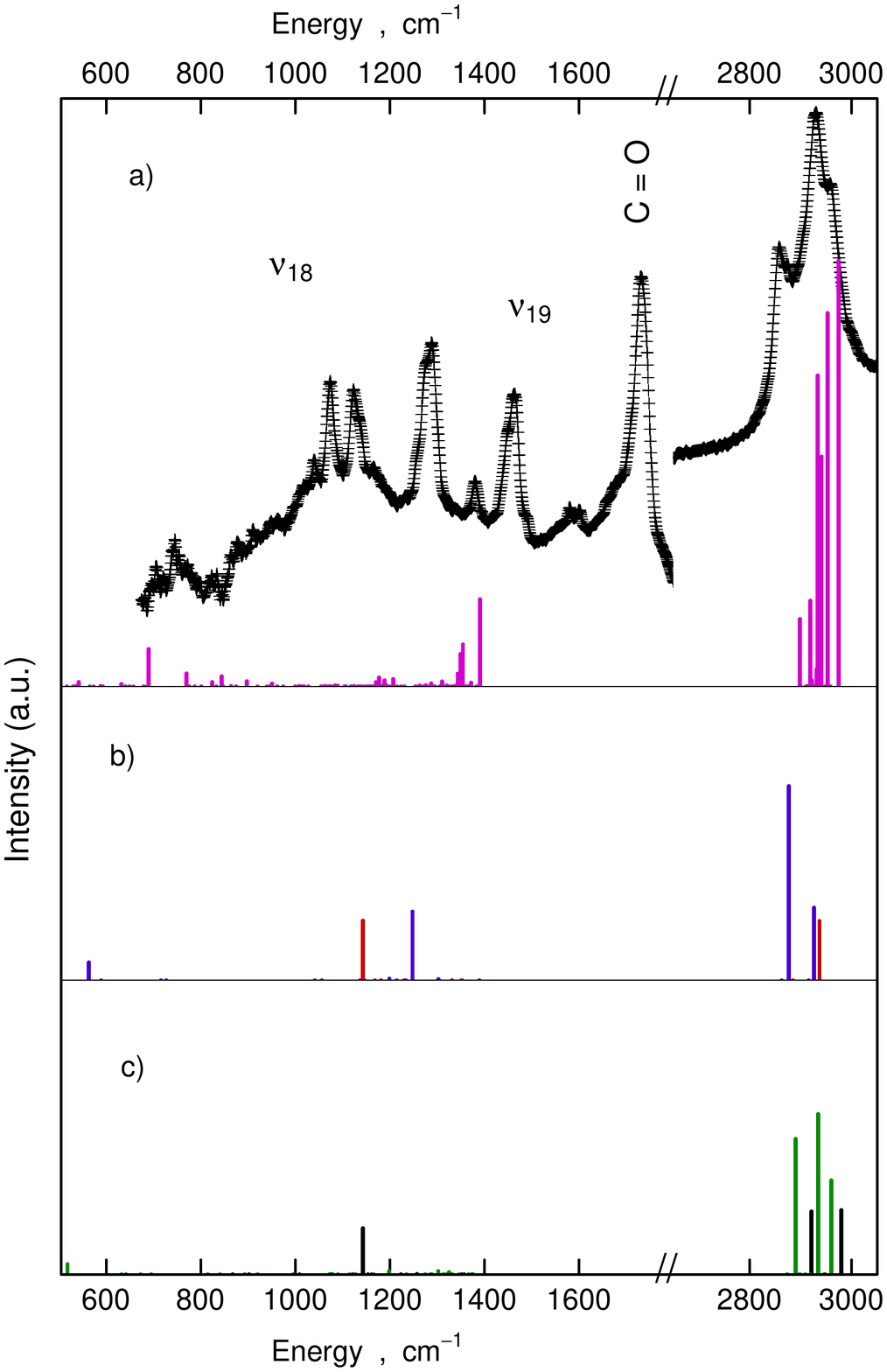}
\caption{Experimental and computed infrared spectra of several graphanes. a) -- experimentally observed IR spectrum ($+$) in comparison to the intensity of IR-active modes of 3-cycle-2-step graphane (vertical bars). b) and c) computed intensity of IR-active modes in the four energetically favorable forms of graphanes with diamond- and lonsdaleite-type of C-C bond respectively. Red, blue, black and green colors stand for A, B, C and D graphane sheets correspondingly \cite{wen:pnas11}.}
\label{f1}
\end{figure}

\begin{figure*}
\begin{overpic}[width = 0.5\textwidth]{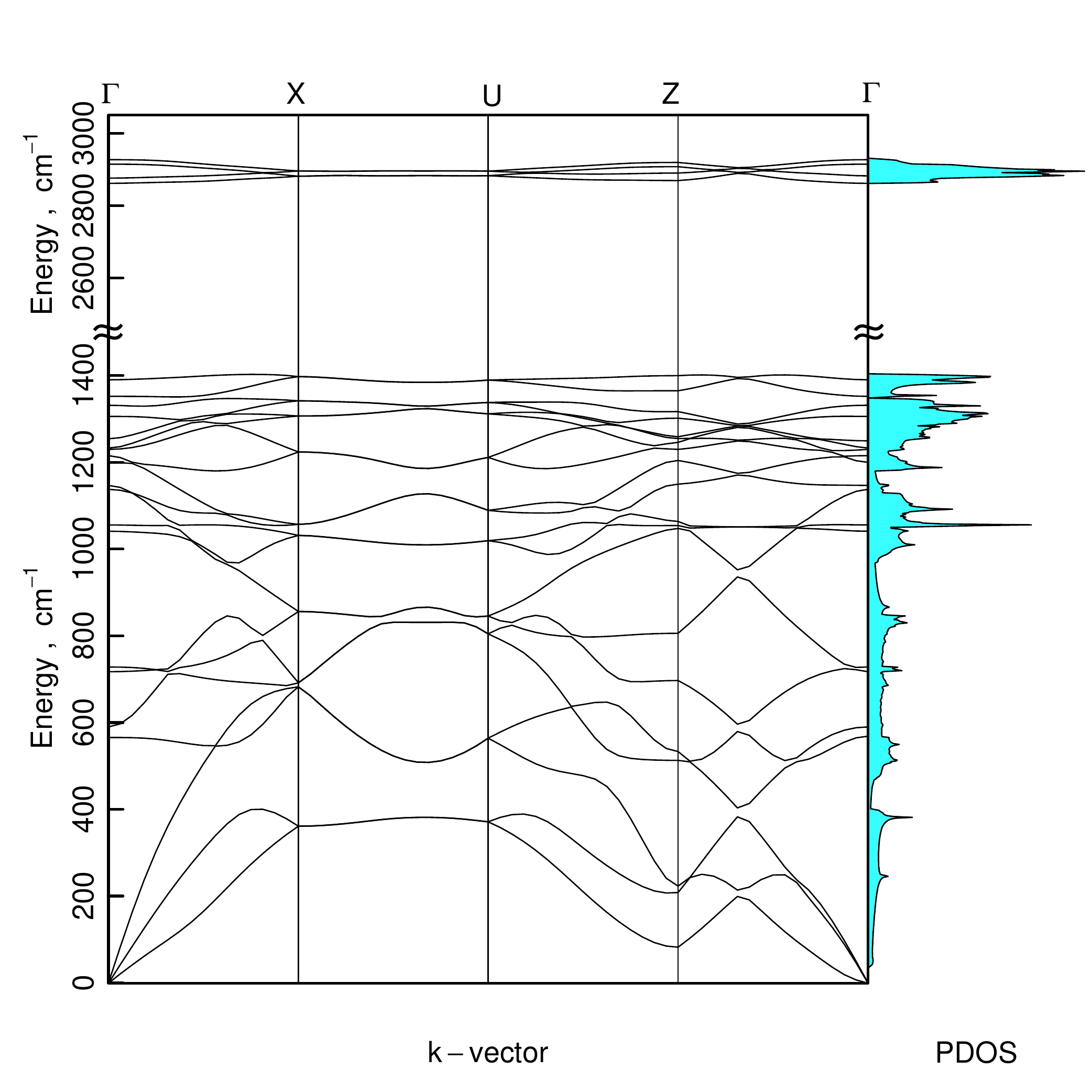}
\put(90,90){a)}
\end{overpic}
\begin{overpic}[width = 0.5\textwidth]{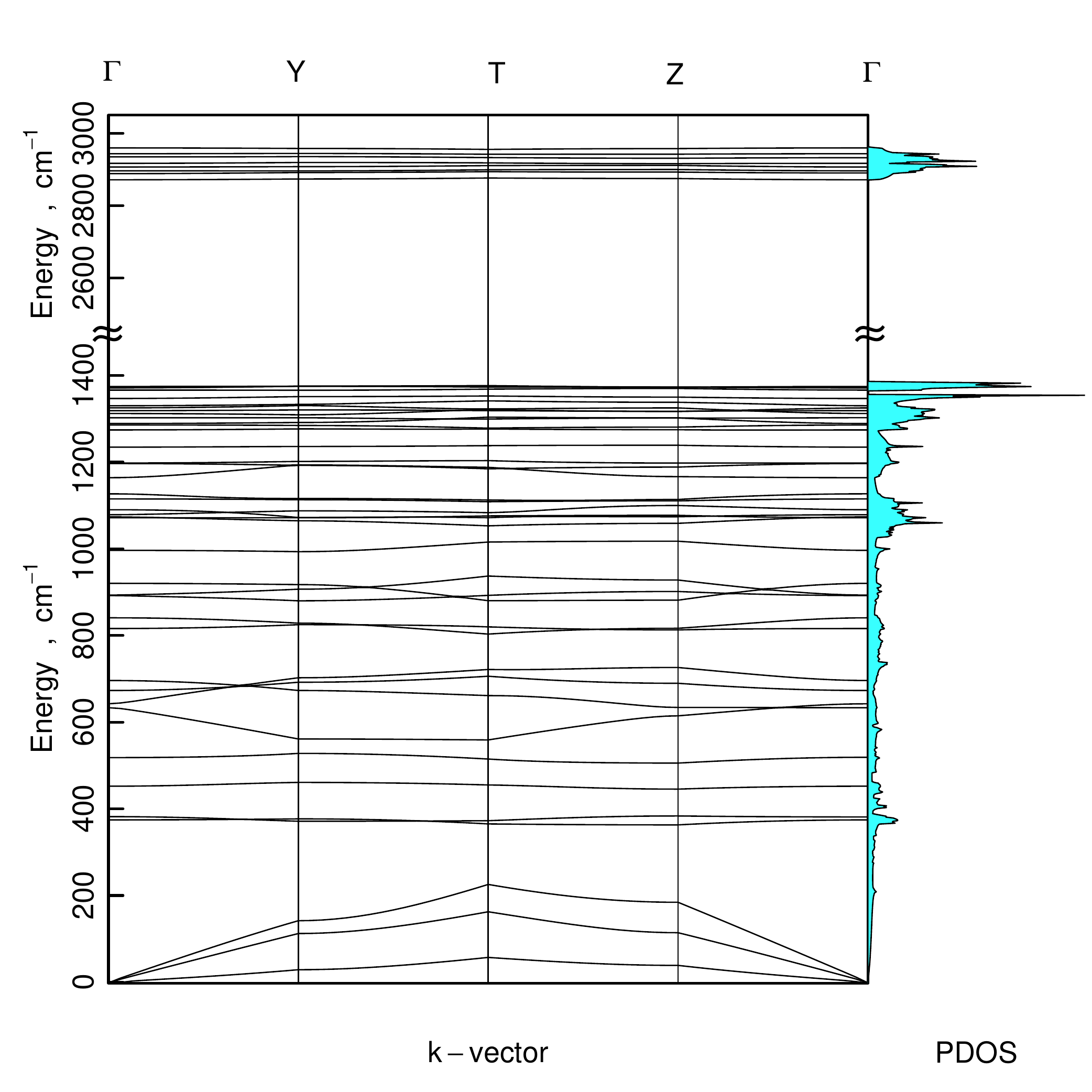}
\put(90,90){b)}
\end{overpic}
\caption{Phonon dispersion in the two most dense graphane sheets and phonon density of states. a) B-graphane (can be compared with the results of \cite{cadelano:prb10} where it was named washboard or W-graphane), b) D-graphane. }
\label{f2}
\end{figure*}

Despite rather long history of graphane's theoretical studies, the predicted IR spectra were not yet published. For some graphane sheets \cite{sofo:prb07,cadelano:prb10,peelaers:apl10} the density of states of phonon modes in $\Gamma$ point and dispersion spectra of phonon modes were calculated, which provide energies of IR-active modes, but not their intensities. In Fig.~\ref{f1} we show relative intensity of IR-active phonon modes obtained for all four simple graphanes and one hybrid structure. These intensities are compared with already published experimental data \cite{kondrin:cec17}. Correspondence between our calculations and published phonon dispersion curves and phonon density of states of two simple graphane sheets can be judged from data shown in Fig.~\ref{f2}.

Calculated vibrational modes in simple graphanes can be divided into three groups. The most intense set of IR-active modes (observed in all graphanes) corresponds to stretching of C-H bonds and can be further sub-classified into symmetrical and asymmetrical oscillations of hydrogen atoms with slightly different energies (in the range of $\approx$ 50 cm $^{-1}$). This attribution, according to work of Dischler {\em et al} \cite{dischler:ssc83} was already done in Refs.~\cite{antonov:c16,kondrin:cec17}. However, our calculations demonstrate that oscillations of hydrogen atom are sensitive to the type of C-C bond of carbon to which  hydrogen  is attached. The presence of lonsdaleite-type C-C bond should additionally enhance this splitting and ensure appearance of stretching mode with energies above 2950 cm$^{-1}$. The presence of IR-modes of this energy was already observed in experiments\cite{antonov:c16,kondrin:cec17}. Variation of intensity of modes with energy above 2950 cm$^{-1}$ reported in these works \cite{antonov:c16,kondrin:cec17} might indicate different concentration of lonsdaleite-type bonds in products obtained by two different methods. In any case, splitting of stretching C-H mode would serve a strong indication of non-equivalence of corresponding moieties (sp$^3$ hybridized CH group) in synthesized crystal structures.

\begin{figure*}
\begin{overpic}[width = 0.5\textwidth]{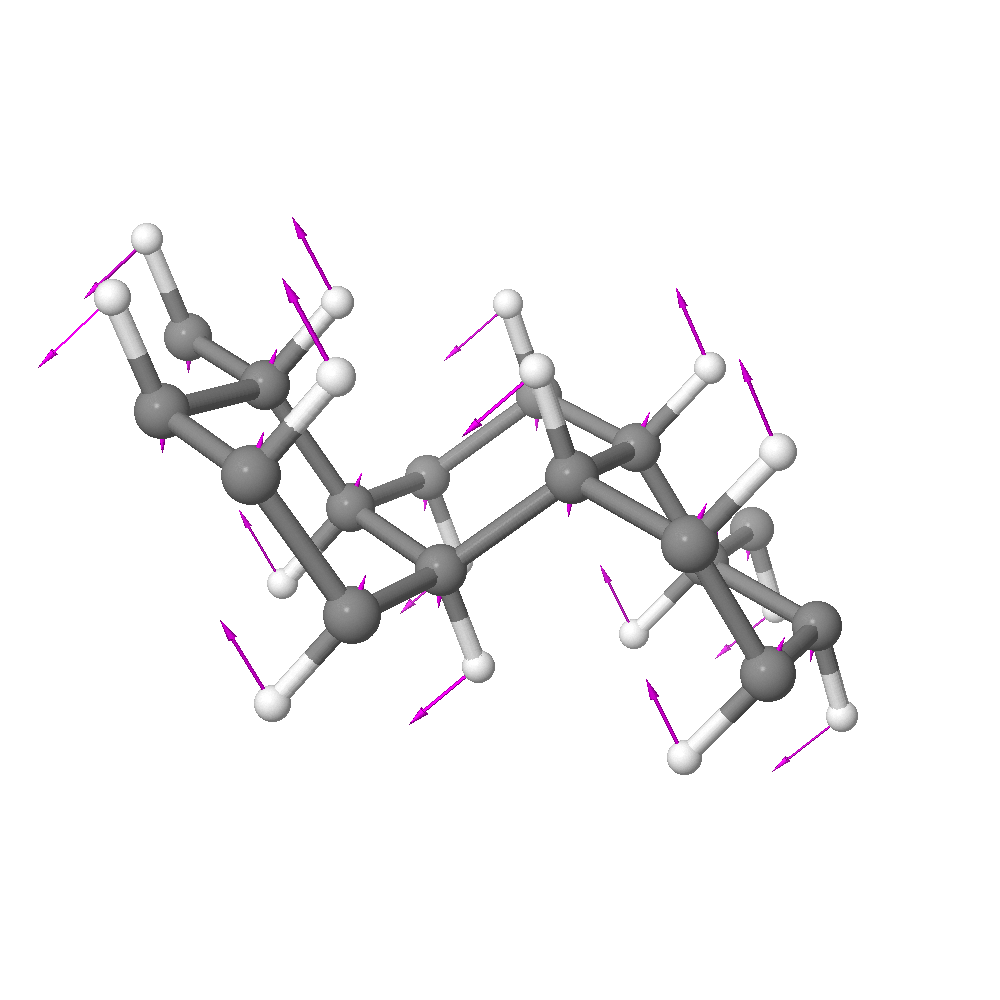}
\put(90,90){a)}
\end{overpic}
\begin{overpic}[width = 0.5\textwidth]{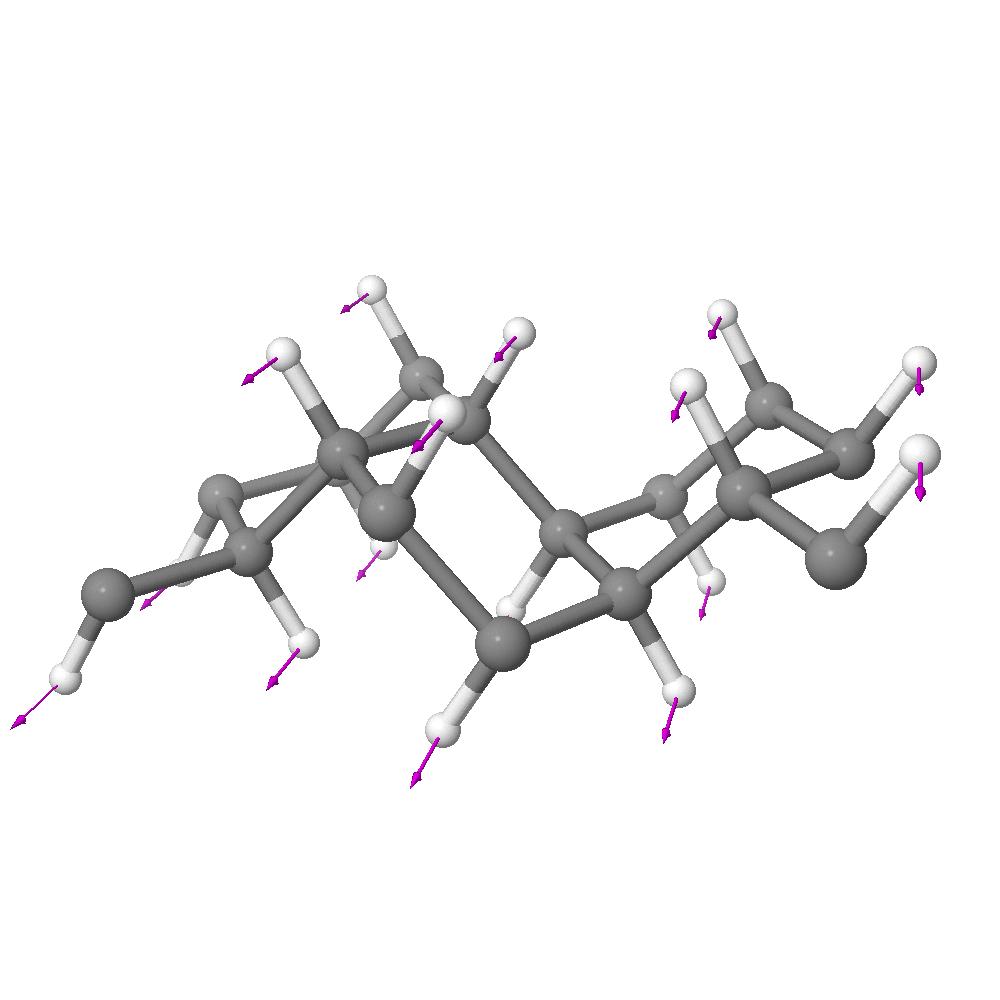}
\put(90,90){b)}
\end{overpic}
\caption{ Two IR-active C-H bending vibrational modes observable in B-graphane. Panels a) and b) correspond to modes with energies 563.56 and 1248.28 cm$^{-1}$ respectively.}
\label{vibr}
\end{figure*}

Observation of additional absorption peaks in low energy regions is even more interesting. The range 1500-2700 cm $^{-1}$ for pure alifatic C-H compounds is forbidden.  A prominent peak at 1730 cm $^{-1}$ (Fig.~\ref{f1}) is probably caused by C=O oscillations and ultimately connected to oxygen contamination in the process of synthesis  or subsequent oxidation. Two other narrow peaks at 1460 and 1040 cm $^{-1}$ can be ascribed to the most intense modes ($\nu_{18}$ and $\nu_{19}$)\cite{ciabini:jcp02} of unpolymerized benzene molecules. Still, there is numerous set of absorption peaks in the lower energy range $300-13500$ cm $^{-1}$, which, in our opinion, should be ascribed to various bending modes of C-H groups.  The observed set of modes can not be ascribed to any of the four simple graphane structures, and this can serve as a collateral indication of complexity of crystal structure of recovered polymeric product. Moreover, symmetric bending mode contributing to the spectral density above 1000 cm$^{-1}$ strongly depends on the crystal structure of graphane sheets. For graphane sheet D it is practically absent, while for 3 other its energy and intensity vary (Fig.~\ref{vibr}). For complex hybrid structure, this mode reaches the highest energy.

Additional asymmetric bending C-H modes contribute to absorption bands below 1000 cm$^{-1}$. This oscillation can be regarded as wavy vibration of graphane sheet (Fig.~\ref{vibr}).  A-graphane has no IR-active vibrational modes in this range, and all other simple graphanes have one or two weak absorption peaks in the frequency range 350-650 cm$^{-1}$. So, the presence of numerous set of absorption peaks can be ascribed to none of the simple graphane structures. It may indicate significantly more complicated structure and larger unit cell of the obtained C-H polymeric crystals.

\section{Synthesis of 3D hydrocarbons by copolymerization of benzene and acetylene}

\begin{figure}
\includegraphics[width = \columnwidth]{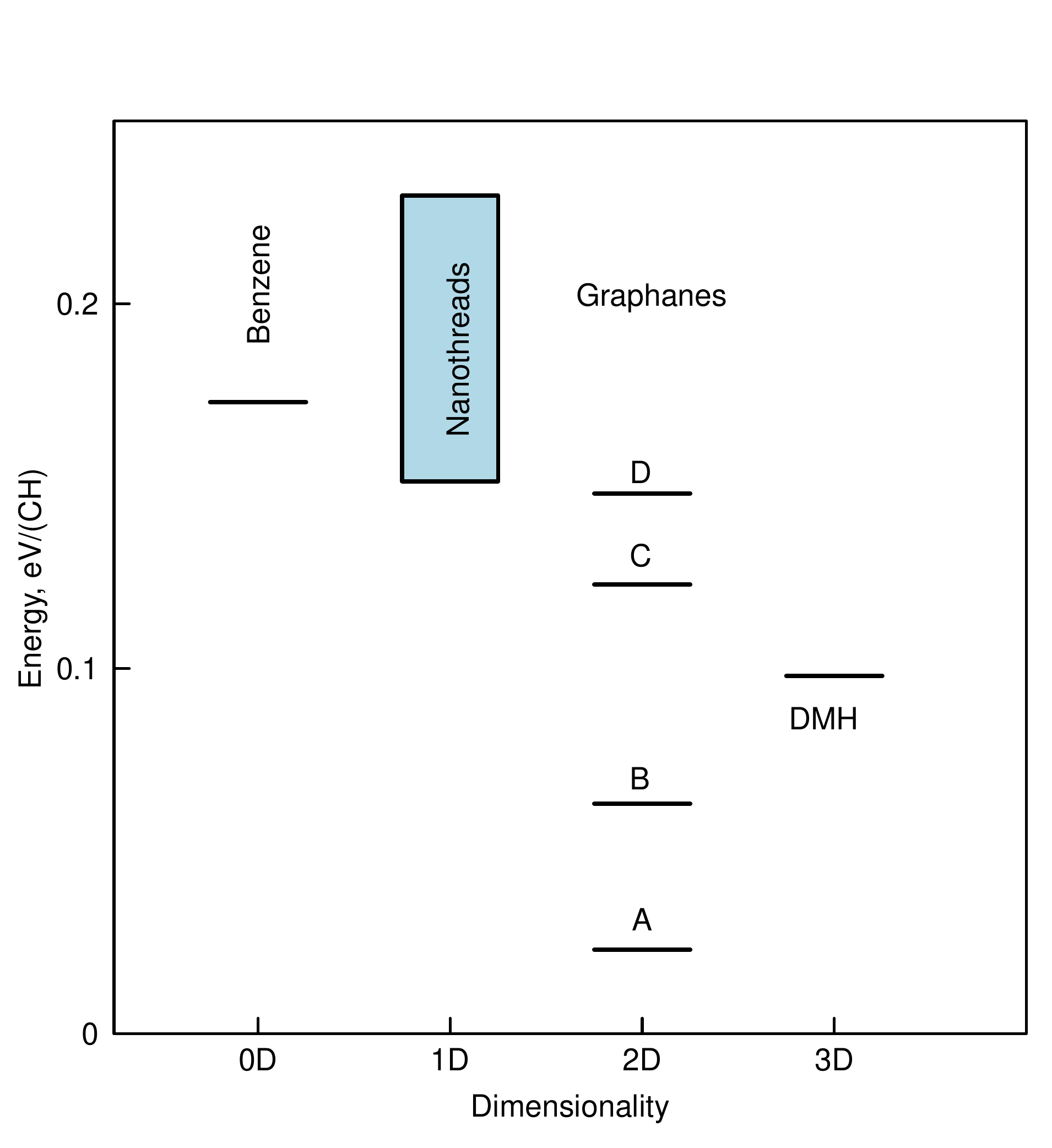}
\caption{Relative stability of several hydrocarbon polymers versus dimensionality of covalent network. Calculated data from Refs.~\cite{wen:pnas11,chen:jacs15,kondrin:acb16} were used. Shaded areas demonstrate the spread of energies of closely related crystal structures.}
\label{diag}
\end{figure}

The energy of hypothetical 3-dimensional DMH structure proposed before \cite{kondrin:pccp15,kondrin:acb16} is not differ uch from pure graphanes' (see energy diagram in Fig.~\ref{diag}) and well below the energy of benzene. However, DMH-like covalent network can not be obtained by polymerization of pure benzene, because the synthesis of its building block requires either including additional components or cracking benzene ring in the process of synthesis. It was proposed earlier \cite{kondrin:acb16} to synthesize this crystal structure  by copolymerization of benzene and acetylene in proportion of 2:1. Acetylene is poorly soluble in benzene and prone to oligomerization, so both these factors have to be taken into account in practical synthesis and determination of required ratio of initial components. However, attempts for copolymerization of benzene and acetylene with 1:1 ratio was carried out recently \cite{huang17} by high-pressure synthesis at room temperature. 

The data obtained by transmission electron microscopy during pressure release were considered \cite{huang17,ward17} as evidence of formation of {\em bcc} carbon phase similar to $\gamma$ silicon with relevant scaling of bond length \cite{johnston:jacs89,nemeth:sr15}. This phase, also known as $i$-carbon or BC8 phase, is believed to be thermodynamically stable at pressures around 1 TPa \cite{correa:pnas06}. The authors implicitly assume that during compression, decomposition of hydrocarbon phases occurs, and molecular hydrogen, formed in the process, leaves the diamond anvil cell during subsequent pressure release. However, BC8 phase is the densest of the carbon phases (which explains its thermodynamic stability at extremely high pressures), so it hardly can produce such a rich set of diffraction peaks observed in Ref.~\cite{huang17}. Moreover, to decrease the crystal symmetry from $Ia\overline{3}$ ($\gamma$ silicon) to $P 2_1 3$ (proposed in Ref.~\cite{huang17}) the crystal structure should be significantly distorted. Such a distortion can be achieved by ``punching holes'' in carbon structure by means of hydrogenation. One of possible structures with 1:1 C:H ratio  and  similar symmetry is K4 hydrocarbon, theoretically described earlier \cite{lian:jcp13}. Its carbon network resembles polymeric nitrogen ($I 2_1 3$ space group) predicted and synthesized some times ago \cite{mailhiot:prb92,eremets:nm04}. Covalent C-C network in K4-hydrocarbon can also be regarded as a result of ``halving'' of parent BC8 carbon structure in the same sense as DMH structure can be regarded as a half of diamond's one. Despite geometric attractiveness of this crystal structure  (it fully deserves its name of ``3-coordinated diamond''\cite{hyde:anie08}), there is no P-T domain where it would be thermodynamically stable. Some other hypothetical polymeric C-H structures significantly outperform it. Another obstacle for practical synthesis of this structure is that the smallest carbon ring found there is a cycle of 10 carbon atoms. Therefore, its synthesis from hydrocarbons mixture consisting of benzene and acetylene would require at least breaking of benzene rings.

\begin{figure}
\includegraphics[width = \columnwidth]{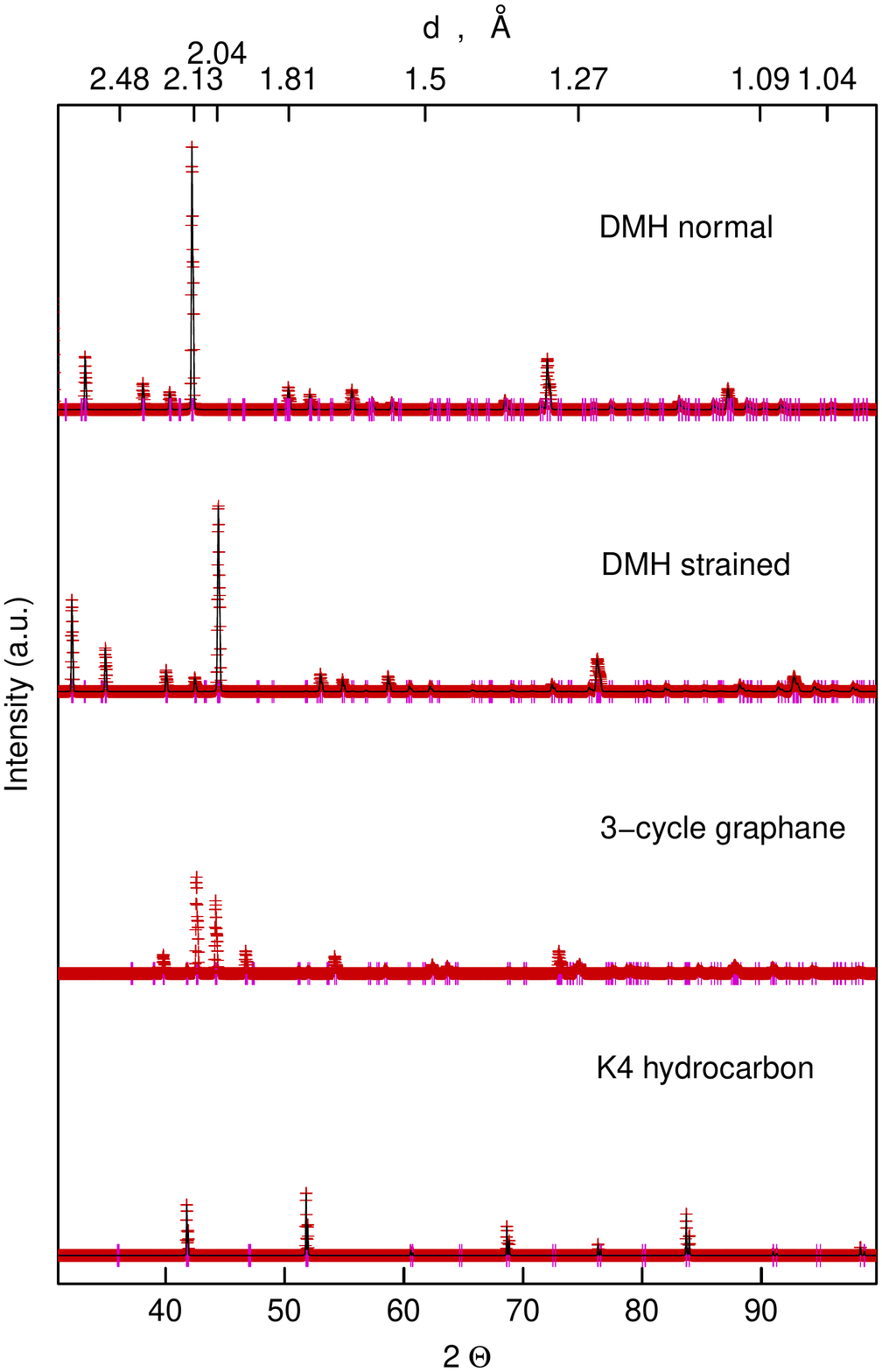}
\caption{Simulated diffraction patterns of several hydrocarbons (see text for details), assumed to match the crystal structure of benzene and acetylene copolymers. Top axis scale provides interatomic distances, according to TEM experimental data \cite{huang17}.}
\label{xray}
\end{figure}

Still, we believe that experimental data provided in Ref.~\cite{huang17,ward17} might indicate presence of polymeric hydrocarbon structures.  It should be noted that hydrogen (the most common isotope) is practically invisible to diffraction techniques including TEM, XRay and neutrons, so crystallographic sites occupied by it in practice can be taken for voids. Comparison of observed and simulated diffraction patterns of several hydrocarbons is depicted in Fig.~\ref{xray}.  The most intense peaks of crystalline hydrocarbons are in the low angles ($d > 2.5$\AA\ corresponding to rather large voids present in these structures), which is inaccessible to the electron diffraction technique. Here  we compare only the regions where experimental data is available (Fig.~\ref{xray}). Since electron microscopy images \cite{huang17} suggest cubic symmetry of obtained crystals, we consider only those structures, which can be regarded as a subset of cubic one, including layered 3-cycle graphane \cite{he:pssr12}.

Our preliminary conclusion is that the most probable structure is the compressed DMH \cite{supplementary}, which at least reproduces positions of some observed diffraction peaks. The contraction by about 4 \% of C-C bond length in comparison to relaxed DMH structure obtained by the DFT calculations at ambient pressure \cite{kondrin:pccp15,kondrin:acb16} might be explained by presence of additional strain, caused by embedding of the crystal grains of this phase into amorphous hydrocarbon matrix, formed under high pressure conditions. However, to achieve a final conclusion, further investigation of relative peak intensities,  obtained by different diffraction techniques on samples different isotope composition is surely required.

\section{The structures of polymeric hydrogen cyanide }
Pure hydrocarbon structures can be interesting to application as possible energy storage devices\cite{sturala:caej17}. Still, functionalization of polymeric hydrocarbons is very important, because it can endow it with very interesting properties concerning electronic and ionic transport \cite{silveira:pccp17,sturala:caej17}. In practice, the two possibilities should be considered: ``in plane'' substitution of carbon atoms in the C-C bonded sheet and chemisorbtion or substitution of hydrogen (out of plane atoms) by some other functional groups. Both these strategies can be important from practical point of view, but require different synthesis processes. 

\begin{figure}
\includegraphics[width = 0.8\columnwidth]{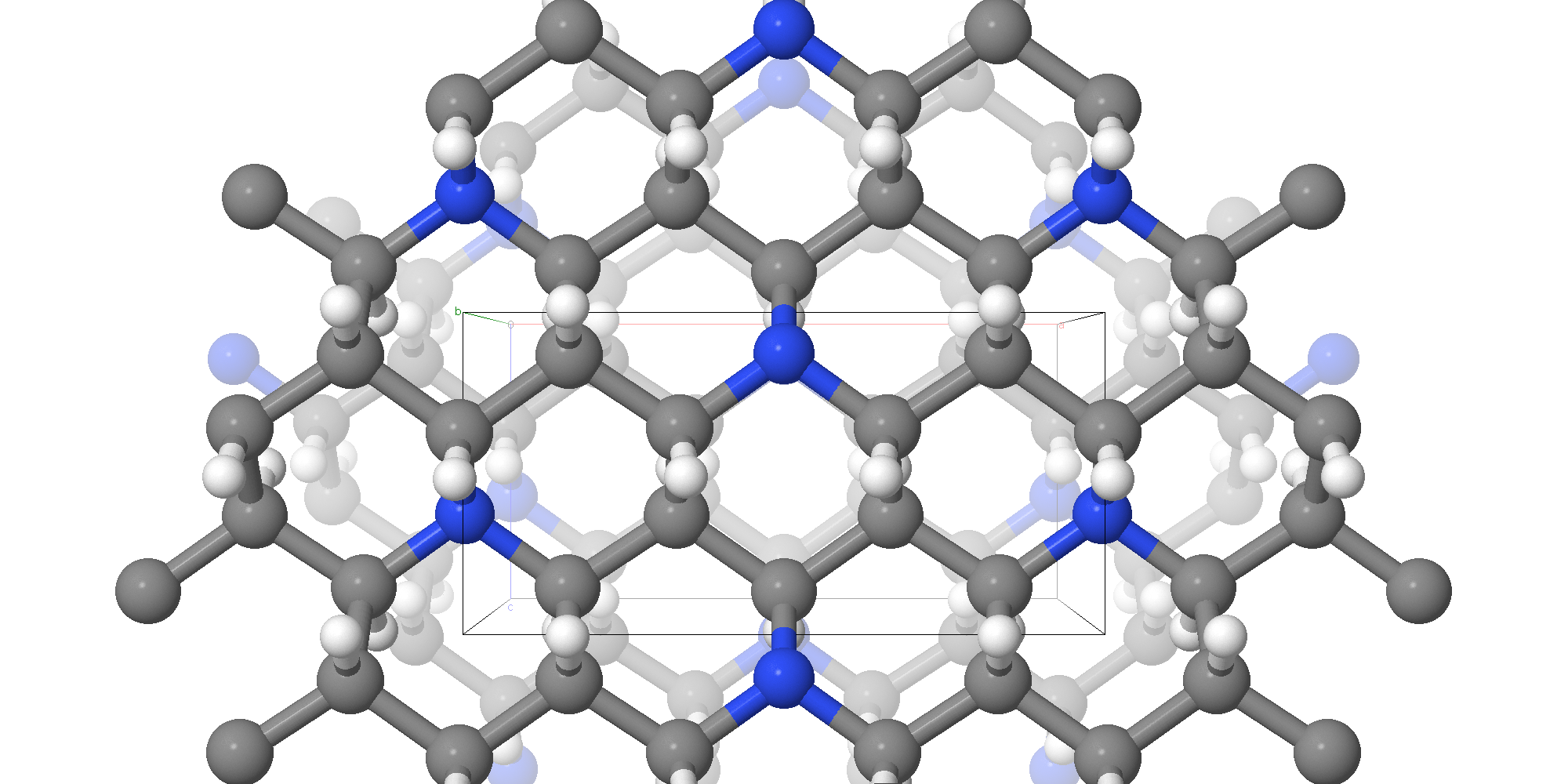}
\caption{One of the polymerized pyridine structures.}
\label{poly-pyr}
\end{figure}

The first attempt of such synthesis was carried out by polymerization of pyridine (C$_5$H$_5$N) under HPHT conditions \cite{kondrin:cec17}. Diffraction pattern of recovered product turned out to be very similar to that of pure hydrocarbon, synthesized under same pressures. It clearly indicates the layered structure of obtained structure, but with slightly different structural parameters, compared to pure hydrocarbon graphane. Still, description of obtained mixed hydrocarbon-nitrogen structure is difficult. Interplay between  corrugation of graphane sheets  and distribution of nitrogen atoms produces very diverse set of possible structures. For the structures based on B-type graphane, at least two different polymeric pyridine structures can be proposed \cite{supplementary}. Despite the large  difference between C-C and  C-N lengths (1.54 \AA\ and 1.47 \AA\ respectively), which would be expected to produce large strain in the polymerized covalent network, the energies of both polymeric structures turn out to be lower than the energy of molecular pyridine (by about 0.32 eV per (CH)$_5$N molecule). The structure of the most energetically favorable polymeric pyridine phase  (with $Pmn2_1$ space group) is depicted in Fig.~\ref{poly-pyr}. Interesting to note that relative shortness of C-N bond in case of polymeric pyridine results in contraction by about 1\%-2\% of in-plane lattice parameters in comparison to the corresponding purely hydrocarbon structures.  

\begin{figure*}
\begin{overpic}[width = 0.32\textwidth]{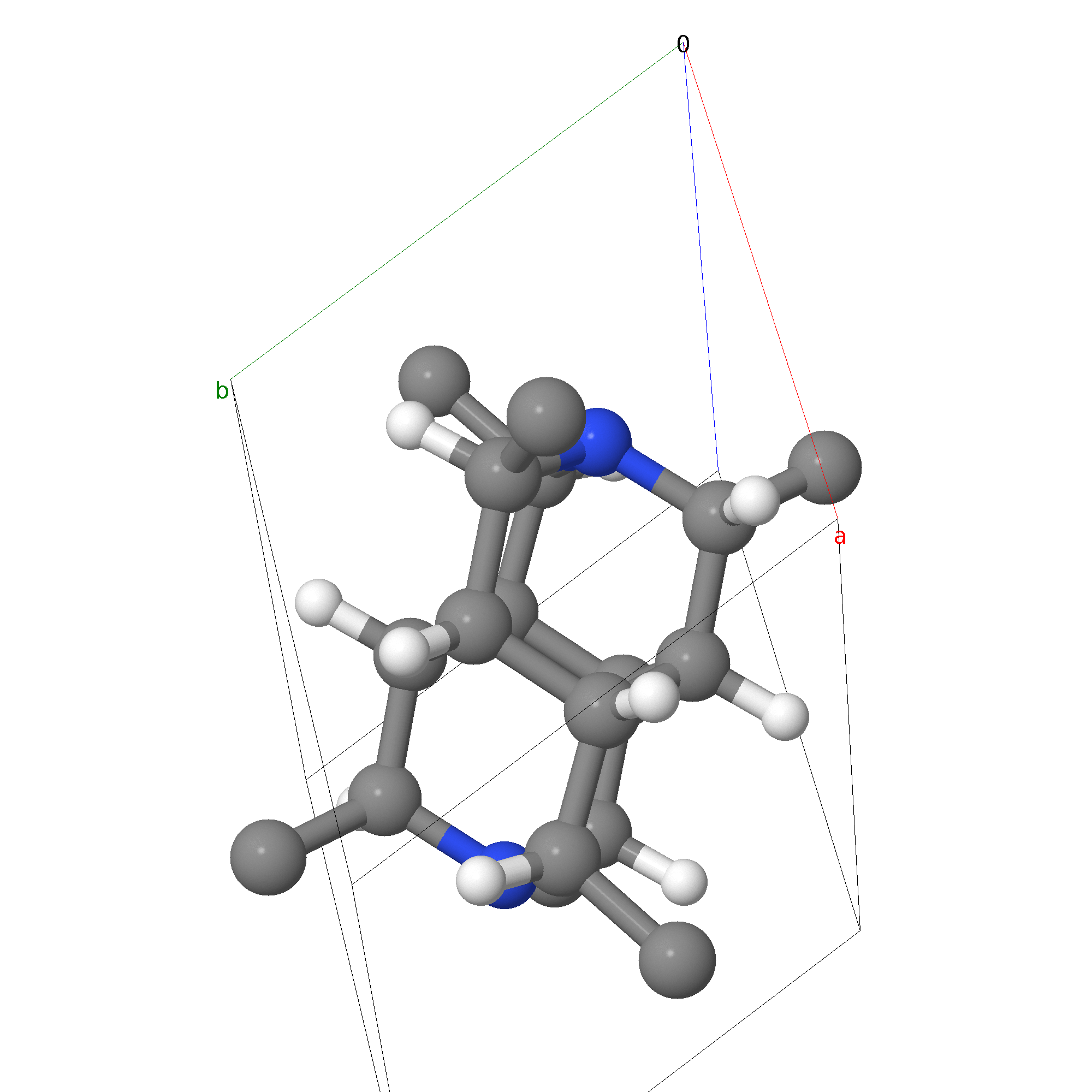}
\put(90,90){a)}
\end{overpic}
\begin{overpic}[width = 0.32\textwidth]{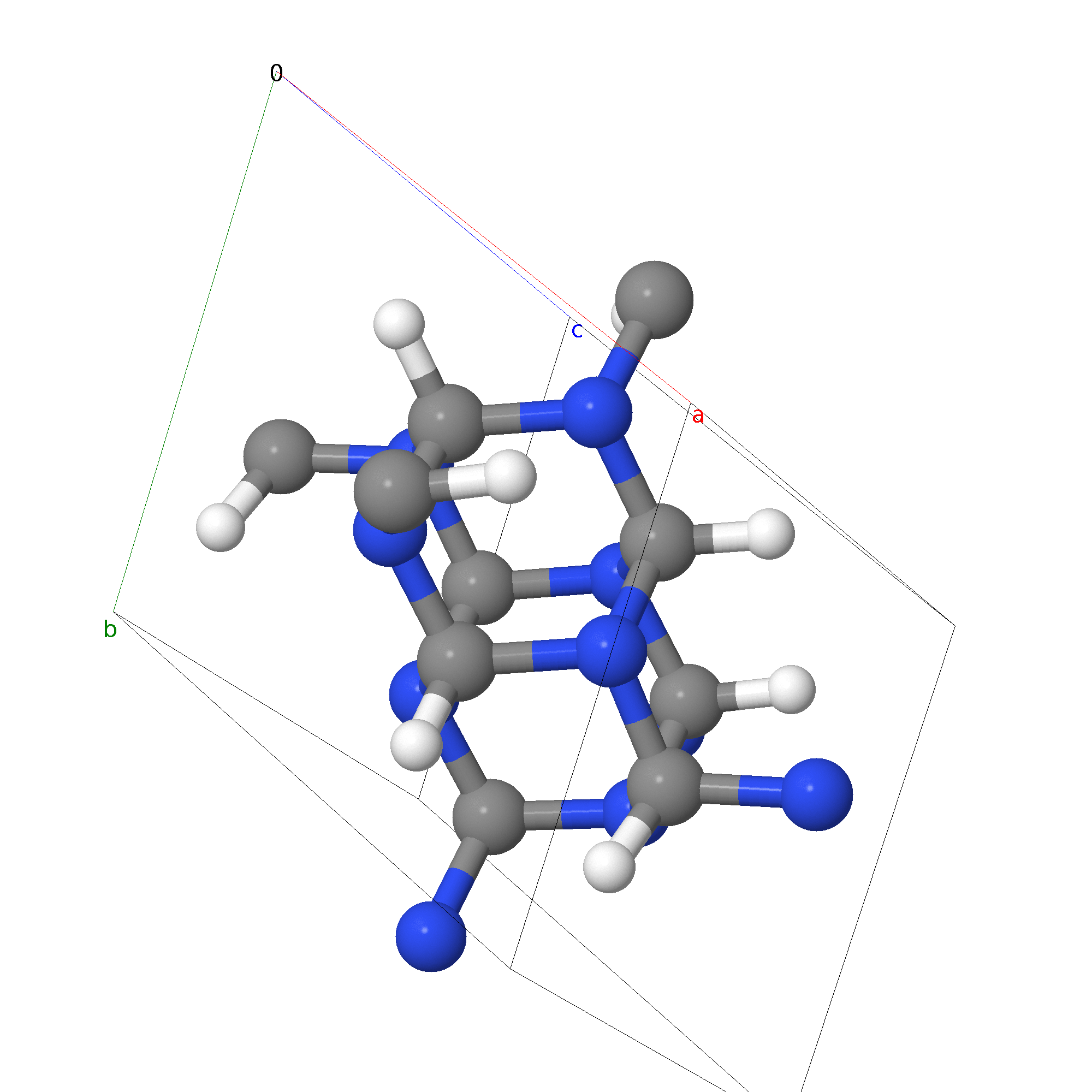}
\put(90,90){b)}
\end{overpic}
\begin{overpic}[width = 0.32\textwidth]{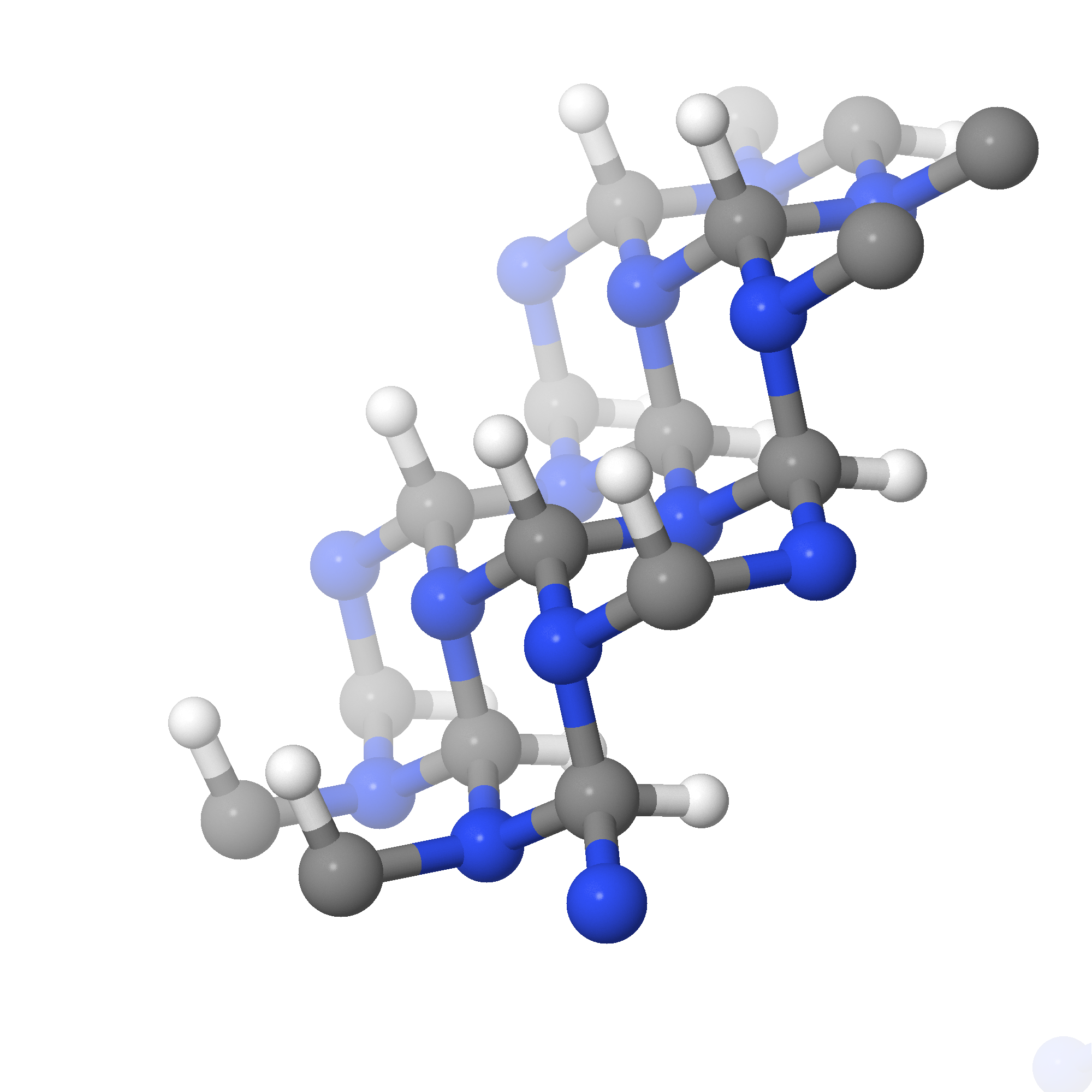}
\put(90,10){c)}
\end{overpic}
\caption{Several hypothetical structures, which would be obtained by copolymerization of acetylene and pyridine (a), and by polymerization of CNH molecules (b and c). }
\label{cnh}
\end{figure*}

To investigate the influence of nitrogen on polymeric hydrocarbon network, we also considered one 2D (graphane-like) and two 3D crystal structures (DMH-like)\cite{supplementary}, which would be obtained by either copolymerization of pyridine and acetylene (Fig.~\ref{cnh}~a), on the one hand, or by polymerization of pure CNH, on the other (Fig.~\ref{cnh}~b and c). At moderate pressure region (below 100 GPa) any single N-N covalent bonded compound tends to turn into monomeric triple bonded molecular N$_2$, this also imposes additional restrictions on possible crystal structures. So, we consider only hydrogen cyanide polymers where the nitrogen atom is bound only to carbon. So the structures obtained by polymerization of CHN molecules are binary (composed by intermixed N and C-H groups) and they lacking inversion center (see Fig.~\ref{cnh}~b and c). Thus, their symmetry is inferior in comparison to  corresponding polymeric hydrocarbon one. Having in mind polymerization under high pressure conditions, we considered only the most dense conformation of CNH ``graphane'' (corresponding to the B-type purely hydrocarbon graphane sheet) (Fig.~\ref{cnh}~c).

Theoretical investigation of copolymerization of pyridine and acetylene and pure CNH generally reproduces the trends observed in pure hydrocarbons. Pure CNH, as well as acetylene, are rather unstable, so for estimation of energetic stability of proposed polymeric phases we compared the energies of (CH)$_2$ and CNH blocks to the energy  of corresponding links in either benzene or pyridine molecules. All three polymeric CNH structures turn out to be more energetically favorable, than corresponding mixture of benzene and pyridine. 2D polymerized hydrogen cyanide is approximately by 26 meV per CNH group, more energetically favorable than its 3D counterpart. Still, the energy of CNH group in 3D-polymer is about 43 meV per CNH group, which is more energetically favorable than the same group in the pyridine ring. The energy of pyridine-acetylene cocrystal unitcell can be compared with the energies of two pyridine molecules and one third of benzene molecule (C$_2$H$_2$). In this case the energy of polymeric structure is also more energetically favorable than its molecular counterparts and equal to 200 meV per unit cell (C$_{12}$H$_{12}$N$_2$).

We should stress, that resulting crystal structure of polymerized products strongly depends on the ratio between initial components (pyridine, benzene and acetylene). Although direct polymerization of pure cyanide is also possible, it surely involves obvious technical difficulties. 

\section{Conclusions and final remarks}

We conclude that the first-principle calculations and experimental evidence suggest that high-pressure high-temperature treatment of aromatic molecular organics might result in the synthesis of covalently bonded organic polymers. These polymers can provide technological platform for future applications as energy storage and harvesting devices. Proper chemical functionalization would make these materials interesting for optics,  electronic and ionic transport.

\begin{ack}
This work was supported by the Russian Science Foundation (grant no. 14-22-00093). Y.B.L. acknowledges financial support from  the Russian Foundation for Basic Research (grant no. 16-02-01120).  The authors thank Roberto Bini and Matthew Ward for stimulating discussions.
\end{ack}

\section{Appendix: CIF}
\begin{verbatim}
data_3c2s
#3-cycle-2-step graphane

_symmetry_space_group_name_H-M    'P b c m'
_symmetry_Int_Tables_number       57

_cell_length_c                    8.47
_cell_length_a                    10.0
_cell_length_b                    9.203

_cell_angle_alpha                 90.0000
_cell_angle_beta                  90.0000
_cell_angle_gamma                 90.0000

loop_
_atom_site_label
_atom_site_Wyckoff_symbol
_atom_site_fract_z
_atom_site_fract_x
_atom_site_fract_y
H 4g 0.12203130 0.2954317 0.0445837202
H 4g 0.12796870 0.7045660 0.8028417867
C 4g 0.09059866 0.5028002 0.0009054428
C 4g 0.15939886 0.4971907 0.8465171137
C 4g 0.15926150 0.3904578 0.0963273234
C 4g 0.09073864 0.6095479 0.7510990142
H 4g 0.12434570 0.6007204 0.0477938560
H 4g 0.12565934 0.3992847 0.7996186951

data_dmh-strained
#strained DMH structure

_symmetry_space_group_name_H-M "R -3"
_symmetry_Int_Tables_number   148

_cell_length_a   6.6
_cell_length_b   6.6
_cell_length_c  12.24

_cell_angle_alpha   90.000
_cell_angle_beta   90.000
_cell_angle_gamma  120.000

loop_
_atom_site_label
_atom_site_Wyckoff_symbol
_atom_site_fract_x
_atom_site_fract_y
_atom_site_fract_z
C   6c   0.0000000     0.0000000   0.6803534
C   18f  0.2332855     0.0645335   0.5210336
C   18f  0.2362737     0.0599022   0.6415319
H   6c   0.0000000     0.0000000   0.7657836
H   18f  0.3410955     0.2609806   0.3366129
H   18f  0.1104390    -0.2893982   0.5086900


data_poly-pyr_1
#polymerized pyridine (1)

_symmetry_space_group_name_H-M    'P m a 2'
_symmetry_Int_Tables_number       28

_cell_length_a                    7.477000
_cell_length_b                    3.754
_cell_length_c                    11.072


_cell_angle_alpha                 90.0000
_cell_angle_beta                  90.0000
_cell_angle_gamma                 90.0000

loop_
_atom_site_label
_atom_site_Wyckoff_symbol
_atom_site_fract_x
_atom_site_fract_y
_atom_site_fract_z
N1    2c    0.250000000   0.882003608   0.048238601
C1    4d    0.912713132   0.888421815   0.052730563
C2    4d    0.084465372   0.388246741   -0.05021096
C3    2c    0.250000000   0.625450953   -0.05205046
H1    4d    0.905558181   0.745489756   0.139383552
H2    4d    0.086661577   0.241400209   -0.1366115
H3    2c    0.250000000   0.762583688   -0.1414714


data_poly-pyr_2
#polymeric pyridine (2)

_symmetry_space_group_name_H-M    'P m n 21'
_symmetry_Int_Tables_number       31



_cell_length_a                    7.472 
_cell_length_b                    11.072
_cell_length_c                    3.750 

_cell_angle_alpha                 90.0000
_cell_angle_beta                  90.0000
_cell_angle_gamma                 90.0000

loop_
_atom_site_label
_atom_site_Wyckoff_symbol
_atom_site_fract_x
_atom_site_fract_y
_atom_site_fract_z
N   2a     0.000000000   -0.04270883   0.630273102
C   2a     0.000000000   0.057616138   0.373316144
C   4b     0.339684158   0.048645008   0.358898979
C   4b     0.331516540   -0.05431737   0.635698154
H   2a     0.000000000   0.146811554   0.510784833
H   4b     0.351438244   0.134484571   0.506327390
H   4b     0.328517965   -0.1408236   0.490743437


data_copoly_pyridine_acetylene
#copolymeric pyridine-acetylene 3D structure (rhombohedral setting)

_symmetry_space_group_name_H-M    'R-3:R'
_symmetry_Int_Tables_number 148

_cell_length_a 5.791
_cell_length_b 5.791
_cell_length_c 5.791

_cell_angle_alpha 70.0
_cell_angle_beta 70.0
_cell_angle_gamma 70.0

loop_
_atom_site_label
_atom_site_Wyckoff_symbol
_atom_site_fract_x
_atom_site_fract_y
_atom_site_fract_z

N    6c     0.677042119   0.677042119   0.677042119
C    18f    0.450248954   0.347826755   0.765052871
C    18f    0.581266328   0.462843058   0.874454374
H    18f    0.083295554   0.248108393   0.684016452
H    18f    0.801867505   0.081756359   0.62828756

data_poly_NCH_3D
#polymerized HCN in 3D structure (rhombohedral setting)

_symmetry_space_group_name_H-M    'R 3:R'
_symmetry_Int_Tables_number       146

_cell_length_c                    5.481
_cell_length_a                    5.481
_cell_length_b                    5.481

_cell_angle_alpha                 72.4
_cell_angle_beta                  72.4
_cell_angle_gamma                 72.4

loop_
_atom_site_label
_atom_site_Wyckoff_symbol
_atom_site_fract_z
_atom_site_fract_x
_atom_site_fract_y
N     1a   0.685973364   0.685973364   0.685973364
N     3b   0.452847050   0.343011730   0.764638217
N     3b   0.419129334   0.535659676   0.119338079
C     1a   0.317491157   0.317491157   0.317491157
C     3b   0.590174838   0.458999836   0.878147847
C     3b   0.544184372   0.653070815   0.239800161
H     1a   0.226869828   0.226869828   0.226869828
H     3b   0.812037151   0.091829583   0.624740324
H     3b   0.918873113   0.766963421   0.306667838

data_poly_NCH_2D
#polymerized HCN in 2D structure (B-graphane type)

_symmetry_space_group_name_H-M    'P m n 21 '
_symmetry_Int_Tables_number       31

_cell_length_a                    2.383 
_cell_length_b                    11.072
_cell_length_c                    3.580 

_cell_angle_alpha                 90.0000
_cell_angle_beta                  90.0000
_cell_angle_gamma                 90.0000

loop_
_atom_site_label
_atom_site_Wyckoff_symbol
_atom_site_fract_x
_atom_site_fract_y
_atom_site_fract_z
H   2a  0.000000000   0.141344443   0.509196846
C   2a  0.000000000   0.054007593   0.359474707
N   2a  0.000000000   0.951072529   0.629017215
\end{verbatim}

%\bibliography{polymers,pp-remark}

\begin{thebibliography}{10}
\bibitem{supplementary}
{All new crystal structures considered in this paper are described in the
  Crystallograpfic Information File provided in Appendix Section}
\end{thebibliography}
\providecommand{\newblock}{}

\end{document}